\documentclass[a4paper]{article}
\usepackage[utf8]{inputenc}

\usepackage{RR}

\usepackage{hyperref}

\usepackage{listings}
\usepackage{graphicx}
\usepackage{float} 
\usepackage{amsmath} 
\usepackage{amsthm} 
\usepackage{color}
\usepackage{amssymb}

\newtheorem{definition}{Definition}

\RRdate{July 2012}

\RRetitle{File system on CRDT\footnote{This work is a delivrable of french
national research programs ConcRDanT (ANR-10-BLAN-0208).}}
\RRtitle{Les syst\'eme de fichier comme graphe en CRDT\footnote{Ce travail est aussi un délivrable de l'ANR ConcRDanT (ANR-10-BLAN-0208).}}

\RRauthor{Mehdi Ahmed-Nacer \and St\'ephane Martin  \and Pascal Urso}

\RRmotcle{Système distribué, Système de fichier, CRDT, Réplication
  optimiste}

\RRkeyword{Distributed System, CRDT, Optimistic Replication, Data
  Consistency, File system}

\RRresume{ Dans ce rapport nous allons montrer comment gérer une structure
  hiérarchique structuré représentant un système de fichier. Cette structure est
 basé sur la réplication optimiste, chaque utilisateur travaille sur une copie
local et les mises à jour sont envoyées aux autres réplica. Les différentes
répliques observent éventuellement la même vu du système de fichier. Á ce
stade, des conflits entre les mises à jour peuvent avoir lieu.  Nous demandons
l'intervention des utilisateurs pour résoudre les conflits aussi peu que
possible. Dans ce rapport, nous proposons une solution simple et modulaire pour
résoudre ces problèmes en maintenant la cohérence des données.
}
\RRabstract{In this report we show how to manage a distributed
  hierarchical structure representing a file system. This structure is
  optimistically replicated, each user work on his local replica, and
  updates are sent to other replica. The different replicas eventually
  observe same view of file systems. At this stage, conflicts between
  updates are very common. We claim that conflict resolution should
  rely as little as possible on users. In this report we propose a
  simple and modular solution to resolve these problems and maintain
  data consistency.}

  \RRprojets{Concordant}
  \URLorraine

\begin{document}
   \RRNo{8027}
  
  \makeRR

\section{Introduction}
Distributed file systems allows different users to work
collaboratively on large-size projects, such as the collaborative
development on the Linux kernel. When a file system is distributed,
many technical and usage issues should be considered and addressed.
As such issues, we can cite all the issues relative to local file
system plus other due to distribution: network communication, privacy
insurance, distributed access control, fault tolerance, replica
distribution, user coordination, etc.  In this report, we only
consider the problem managing concurrent updates on the file system.

Indeed, when multiple people share and modify the same file system
concurrently, the updates can interfere with each other in such a way
that the file becomes useless and contains conflicts. Some traditional
distributed file systems recommend file locks to ensure that the file
is protected. Unfortunately, this method cannot ensure high
responsiveness for real-time collaboration or disconnected work,
because the initiator of an update should acquire an exclusive
access. On the other hand, optimistic
replication~\cite{saito05optimistic} allows availability, performance
and  supports work in disconnected mode. In an optimistically
replicated file system, data is replicated on each replica, and any
replica can independently modify its own state.  However, optimistic
mechanisms gain this availability by trading off linear consistency.

Anyway, all modifications are sent to other replicas and some
consistency must be ensured. Strong eventual consistency (SEC) ensures
that as soon as replicas have received the same updates, the replicas
host the same data value~\cite{shapiro11conflictfree}. Depending of
approach used, these modifications can be sent as a set of update
operations (aka {\em operation-based}), or sent as a whole new state
(aka {\em state-based}). Most of version control systems (VCS) such as
Subversion~\cite{collinssussman2007SVN} or Git~\cite{git} adopt
state-based approaches, while distributed file systems described in
the literature~\cite{kistler1992coda,richard1992ficus} are mostly
operation-based.

In eventual consistency, since any replica can be updated, two
modifications applied independently may lead to conflicts. For
instance, the addition of a file in a directory conflicts with the
removing of this directory. To maintain a correct hierarchical data
type, a system with optimistic replication must either avoid such
conflicts or recover automatically from them. Conflict-free Replicated
Data Types (CRDT) \cite{shapiro11conflictfree} can be a solution.

This report is structured as fellows. Firstly, we give a state of art
to talk about an existing file systems.  The next section presents an
overview of the Conflict-free Replicated Data Types (CRDT). After
that, we begin a definition of CRDTs generally and describe more
precisely the different solutions to build CRDT based on set
structure. The next section ~\ref{sec:layer} shows a new data
structure based on layers that ensure consistency.
Section~\ref{sec:file_sys} discusses about file system and a conflicts
that can arise in an optimistic replication system, we describe
briefly how conflicts are detected and cover.  The next
section~\ref{sec:file_crdt} describes the solution using CRDTs to
manage conflicts. Finally, we close with a conclusion.

\section{File System}
\label{sec:file_sys}

In this section we present the data type corresponding to a
hierarchical file system. We define the structure of this data type
and the update operations that can be applied on it. Finally we
describe the possible conflicts that can arise in with such a
replicated data structure.

We consider the data structure of file system as a tree containing
elements with a typed content. A content type can be a directory --
that contains other element elements -- or file types. We consider
that file types are automatically detected by the replicated
system. For instance, version control systems consider text and binary
file types.  For shake of simplicity, and as many of heavily used
replicated file systems~\cite{ghemawat2003GFS,kistler1992coda}, an
element in only present once in the tree. I.e., we do not manage soft
or hard links.

\begin{definition}
  A {\em file system element} is couple $(n, c)$ where $n$ is the name
  of the element and $c$ is its content. There exists a function
  $type(c)$ that returns the type of the element according to its
  content. The content of a directory is a collection of elements
  where each name is unique. A {\em file system} is a $root$ directory
  with an empty name.
\end{definition}

We consider the basic operations $add$, $remove$ of files and
directories, and $update$ of files. Operations are defined according
absolute paths.

\begin{definition}
  A path is a list of element names $n.n'.n''.\cdots$. The predicate
  $exists(p, S)$ for a path $p$ and a file system $S$ is defined
  recursively as follow:
\begin{eqnarray*}
 & &  exists(\varnothing, c) = true \\
type(c) = directory \wedge \exists (n',c') \in c & \implies &  exists(n'.p, (n,
c)) = exists(p, c') \\
 else & \implies & exists(n'.p, (n, c)) = false
\end{eqnarray*}
The function $content(p,S)$ returns the content of the element at path
$p$ in $S$. The predicate $prefix(p',p)$ is true if and only if the list
$p'$ is a non-strict prefix of the list $p$.
\end{definition}

The operation $add(p,n,t)$ adds an element with name $n$ and an empty
content of type $t$ under the path $p$. The operation $remove(p)$ (or
$rmv(p)$) deletes the last object (file or directory) of path
$p$.\footnote{We consider the more general case where any path,
  including non-empty directory can be removed.}  Whereas $update(p,
u)$ apply modification $u$ on the file located at path
$p$.\footnote{As some existing distributed file systems or VCS, we may
  restrict the set of operation to only $remove$ and
  $update$. However, for shake of clarity in conflict presentation, we
  kept the $add$ operation.}  We consider that each file content is
managed by a conflict-free replicated data type (CRDT)
\cite{shapiro11conflictfree} corresponding to the type of file.  For
instance text files can be managed using sequence CRDT algorithms such
as WOOT~\cite{oster06data}, Treedoc~\cite{preguica09commutative},
Logoot~\cite{weiss09logoot}, or RGA~\cite{roh11replicated}. Binary
files can be managed using a
Thomas-write-rule~\cite{thomas79rule}. Moreover, any kind of file type
can be managed such as sets, graphs~\cite{shapiro11comprehensive} --
or more usefully -- XML files \cite{martin10scalable}.

%

The usage of all the above operations must follow some pre and post
conditions. The pre and post conditions are local, i.e. they must be
ensured when a local modification is done atomically on a
replica. When applied remotely, if the precondition is not respected,
a conflict occurs. The respect of the post condition depends on how
the conflict is resolved.
\begin{itemize}
\item $pre(add(p, n, t), S) \equiv exists(p, S)$ and
  $type(content(p, S))=directory$ and $\neg exists(p.n, S)$
\item $post(add(p, n, t), S) \equiv exists(p.n, S)$ and
  $type(content(p.n, S))=t$ and $isEmpty(content(p.n, S))$
\item $pre(remove(p), S) \equiv exists(p, S)$
\item $post(remove(p), S) \equiv \neg exists(p, S)$ 
\item $pre(update(p, u), S) \equiv exists(p, S)$ and $u$ is applicable
  on type $content(p, S)$
\item $post(update(p, u), S) \equiv exists(p, S)$ and $content(p, S)' =
  content(p, S) \circ u$
\end{itemize}

With such pre conditions, in case of a concurrent modifications, some
conflicts occurs:
\begin{itemize}
\item $add(p, n, t)||remove(p.n)$ : adding and removing the same element
  concurrently
\item $add(p, n, t)||remove(p')$ with $prefix(p',p)$ : adding an element
  while removing one of its ancestors
\item $add(p, n, t)||add(p, n, t')$ : adding two element with same name
  under same directory
\item $update(p, u)||remove(p')$ with $prefix(p',p)$ : updating an
  element while removing one of its ancestors
\end{itemize}
Contrary to existing distributed file systems we do not consider the
$update(p,u)||update(p,u')$ conflict since file contents are CRDT.
Thus, concurrent updates operation can be applied in any order while
obtaining eventual consistency. The remainder of this section will
discuss the conflicts occurrence in more detail. The next section
describes how we manage these conflicts.
 
Even a simple collaboration of two replica can result in a conflict.
Figure \ref{fig:conflict_add_remove} illustrates this
situation. Directory $Toto$ appears on two replicas. Replica 1 creates
a file $prog.c$ under directory $Toto$ when replica 2 removes
$Toto$. Then, when the replicas merge, the pre-condition of
$add(Toto,"prog.c", t)$ is no-longer true since the directory $Toto$
is not present. This is an $add(a)||remove(b)$ conflict.
  
\begin{figure}[H] 
\centering
\includegraphics[width=10cm]{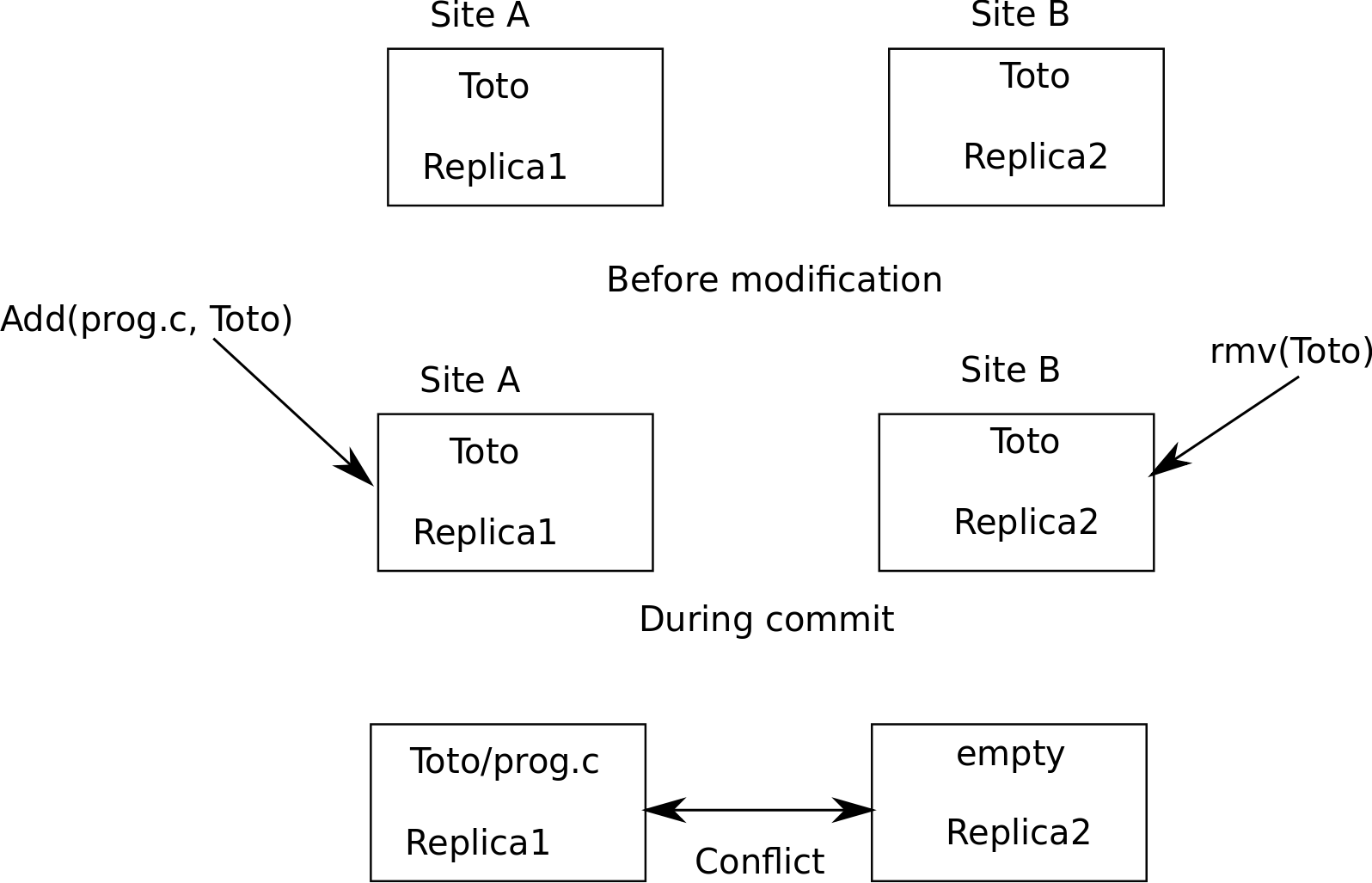} 
\caption{Conflict $add(a)||remove(b)$ }
\label{fig:conflict_add_remove} 
\end{figure} 

\begin{figure}[H] 
\centering
\includegraphics[width=10cm]{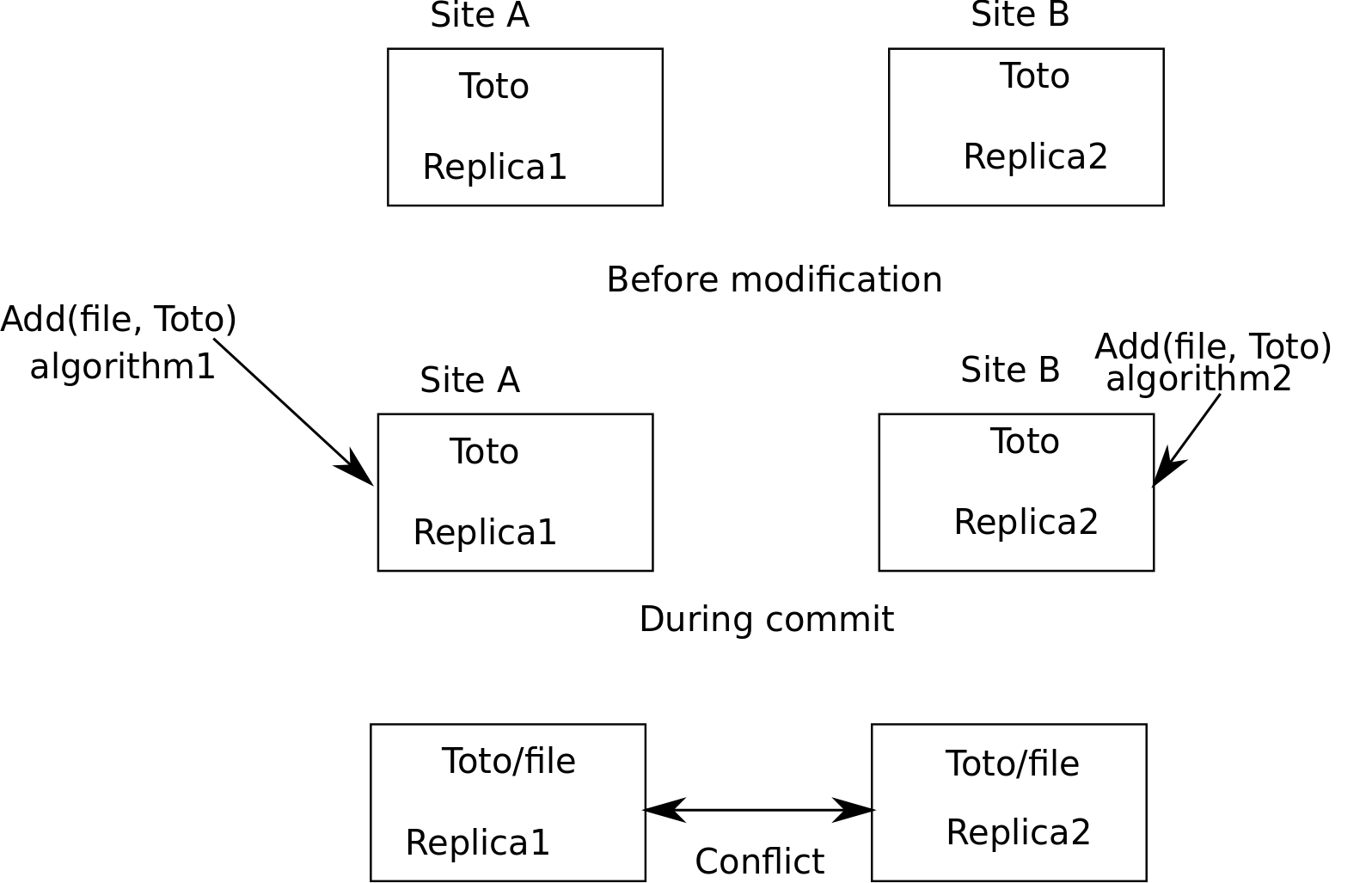} 
\caption{Conflict $add(a)||add(a)$ }
\label{fig:conflict_add_add} 
\end{figure}

Figure \ref{fig:conflict_add_add} illustrates a different kind of
conflict where two users create a same document with same name. In
figure \ref{fig:conflict_add_add} replica 1 creates a document $file$
under directory $Toto$, with the same type. Replica 2 also creates $file$
under directory $Toto$. During the integration of the remote update,
the pre-condition of the second $add$ operation (added path is not
present in the file system) is violated. If the types of the
concurrent $add$ are the same, the system trivially ensures SEC. We
term this $add(a)||add(a)$ conflict \textit{name conflict}.

Another type of conflict is $add(a)||remove(a)$. Indeed, an element
can be deleted and added at the same time.  If replica 1 adds an
element $a$ when replica 2 removes it, divergence occurs.

\begin{figure}[H] 
\centering
\includegraphics[width=5cm]{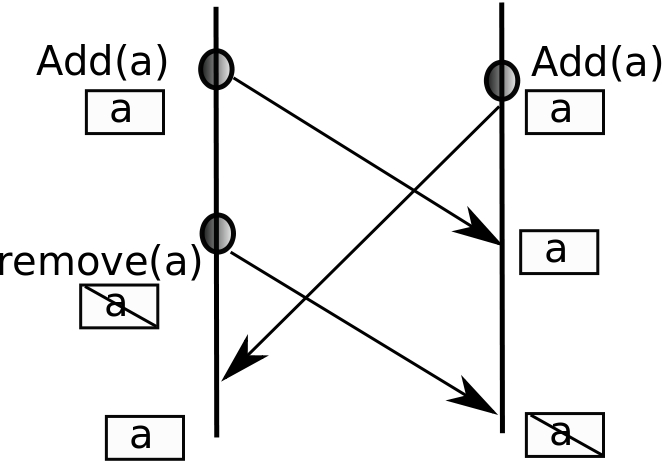} 
\caption{Conflict $add(a)||remove(a)$ }
\label{fig:conflict_add_remove} 
\end{figure}

The last type of conflict is $update(a)||remove(b)$. This conflict occurs
when a replica updates a file content while another removes the file
or a directory in the path to the file. In this case the precondition
of the update operation (the path is present) is violated.

\begin{figure}[H] 
\centering
\includegraphics[width=6cm]{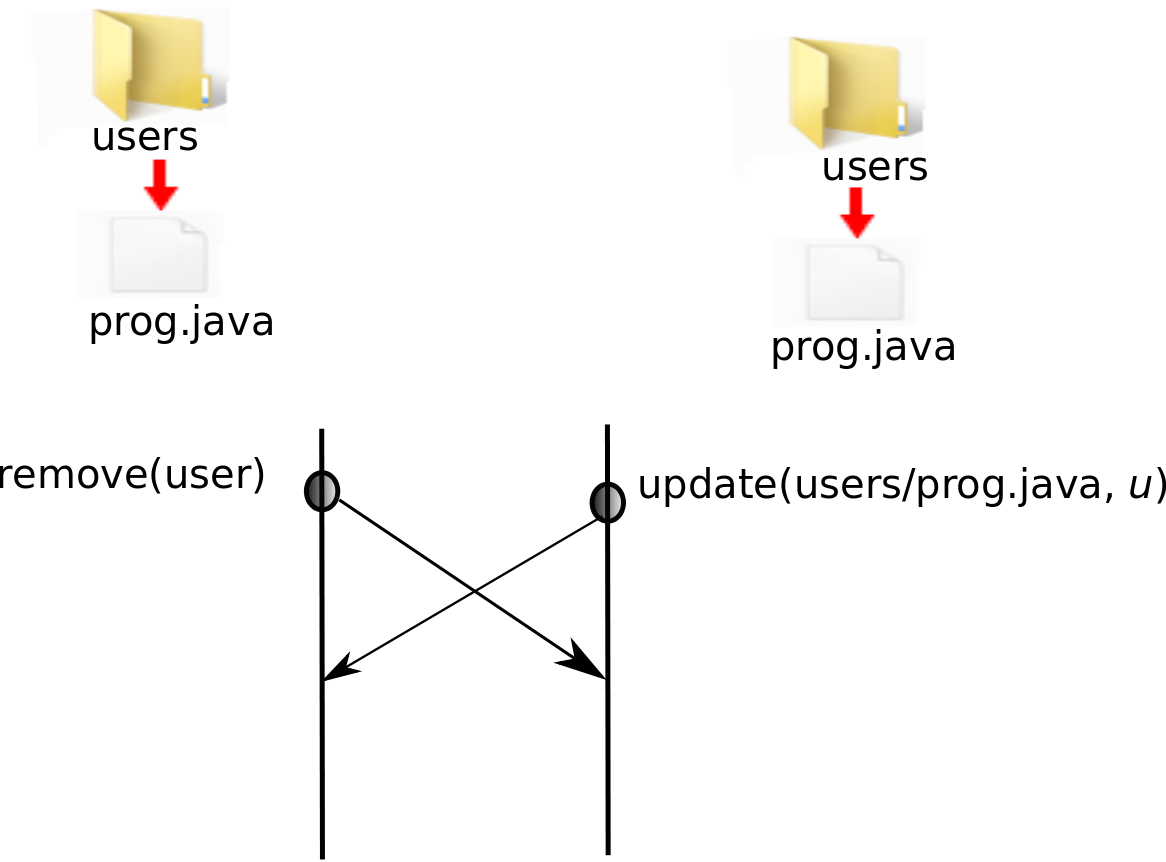} 
\caption{Conflict $update(a)||remove(b)$ }
\label{fig:conflict_update_remove} 
\end{figure}

Our goal is to design a conflict-free replicated data type (CRDT) for
file system. So we need one or more replicated data structure where
such conflicts either cannot occurs or a resolved in a automated
manner. Of course, the obtained data structure must ensure strong eventual
consistency.


\subsection{Ficus}
\label{sec:ficus}

Ficus file system\cite{richard1992ficus} is developped for peer-to-peer
optimistic file replication systems. The conflict possible in Ficus are : \\
- \textit{Update/update conflicts} : It moves the file into a special directory
called an $orphanage$. Each volume has its own orphanage directory located under
its root directory \\
- \textit{Name conflicts} : It occurs when user insert two file with same name
under same directory. Ficus appends unique suffixes to each file name. \\
\textit{Remove/update conflicts} : Ficus allows users to resolve conflict.

Ficus propose also a mechanism to resolve a conflicts automatically except for
name conflicts.

\subsection{Version Control Systems}
Version Control Systems manage files that can be accessed and updated
by multiple user. Today, there are several types of these systems used
such as CVS, SVN, GIT ... ect.  These systems allows multiple users to
work concurrently on a file while ensuring that their work is safe and
not will be lost. Most of these systems are state-based and merge can
be only done manually by a user.  When a user wants to merge
concurrent modification, he obtains a ``best effort merge'' where some
of the conflict (depending on the system) are resolved automatically,
while other have to be resolved by the user before it commits the
merge. The committed merge is a new state in the graph history of the
repository, so no conflict occurs on the repository.

Types of conflicts presented to the user depend on the structure and
data management by the system. For instance, Git~\cite{git}, does not
take into account the directories, it considers the file system as a
hierarchical set of files, while CVS and SVN consider the directories.

\begin{figure}[H] 
\centering
\includegraphics[width=10cm]{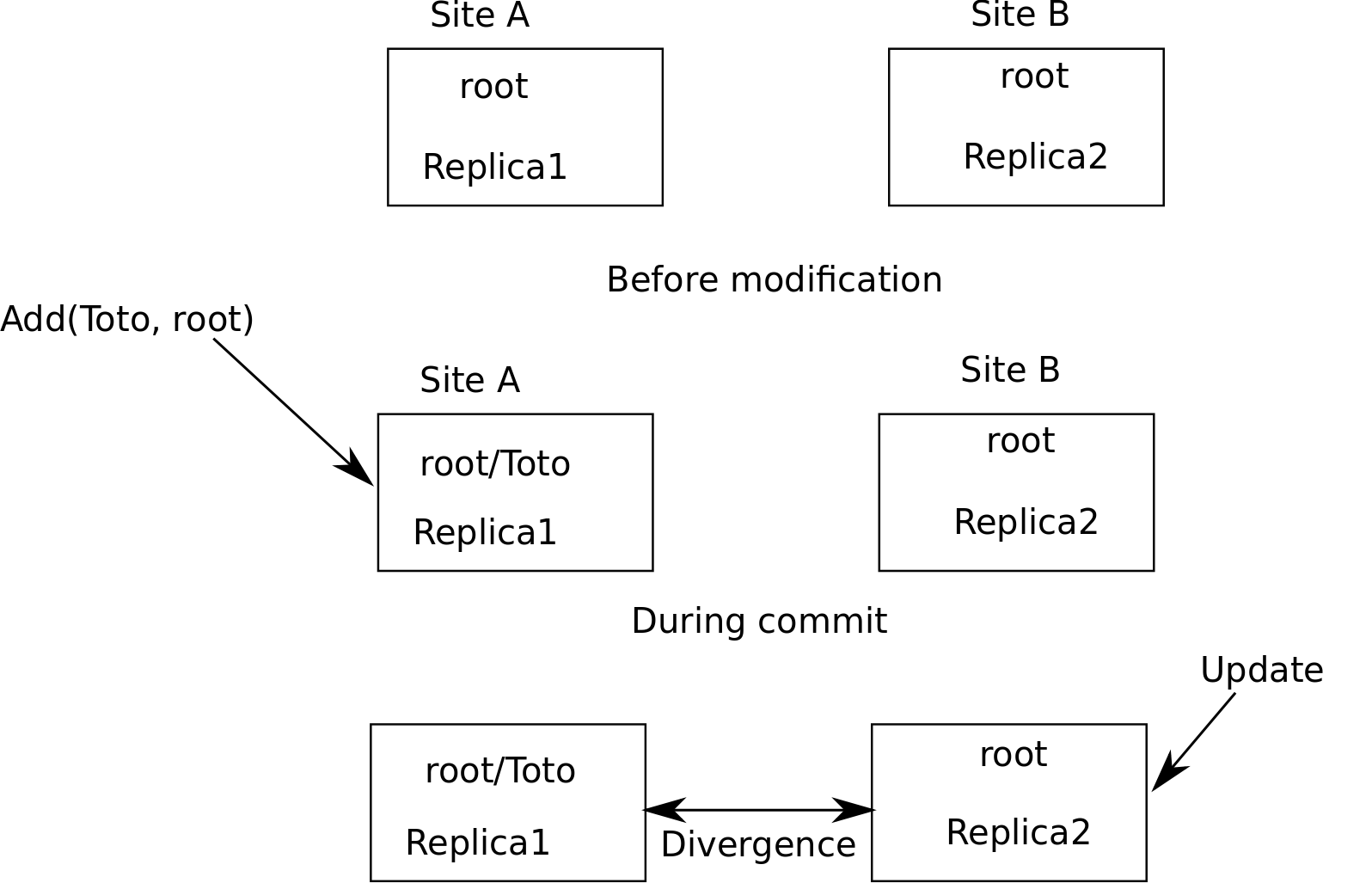} 
\caption{Divergence on git\cite{git}}
\label{fig:divergence_update} 
\end{figure}

On the other hand, a case of divergence can occur depending of data management. In Git~\cite{git},
 the directories are considered only locally. When user create locally
 an empty directory, git does not take it into account in the repository. When users make an update
 as in figure\ref{fig:divergence_update}, two replica may observe a divergence while there are both up-to-date.

\section{Conflict-free Replicated Data Types (CRDT)}

To achieve high responsiveness, data replication is necessary.
When the replicated data is mutable, the consistency between the replicas must
be ensured. The CAP theorem ~\cite{shapiro11conflictfree} 
 states that a replicated system cannot ensure strong Consistency together 
with Availability and Partition tolerance. In many applications, such as collaborative application, where
availability is required by users and partitions are unavoidable, a solution is to allow 
replicas to diverge temporarily and when system is idle, all users observe the same data.

This kind of consistency model is called ``eventual consistency'' which guarantees that if no new update is made to
the object, eventually all accesses will return the same value. The ``strong eventual consistency'' (SEC) model 
guarantees that all accesses return the same value as soon as all update are delivered.  
To ensure SEC, a particular merge procedure is required that 
handles possibly conflicting concurrent modifications.

In what follows you exemplify the CRDT principle by describing some replicated set designs.

\subsection{Set}
\label{sec:set}

 For a CRDT set, we consider two operations: a process can add an element 
 with operation $add(a)$ and can delete it with operation $remove(a)$ . In
a sequential execution, the ``traditional'' definition of the pre- and post-conditions are
 \begin{itemize}
\item $pre(add(a), S) \equiv a \notin S$
\item $post(add(a), S) \equiv a \in S$
\item $pre(remove(a), S) \equiv a \in S$   
\item $post(remove(a), S) \equiv a \notin S$   
\end{itemize}

In case of concurrent updates, the preconditions of $add(a)||remove(a)$
conflict.

\begin{figure}[H] 
  \centering
  \includegraphics[width=8cm]{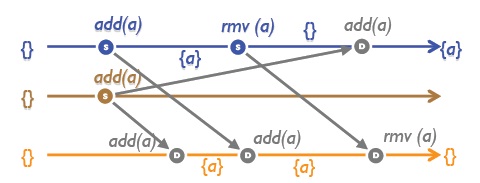}
  \caption{Set with concurrent addition and remove~\cite{shapiro11comprehensive}}
  \label{fig:set}
\end{figure} 

Thus, a set CRDT has different global post-conditions in order to take
into account the concurrent updates while ensuring eventual
consistency. Each CRDT has a {\em payload} which is an internal data
structure not exposed to the client application, and {\em lookup}, a
function on the payload that returns a set to the client
application. For a set CRDT, the pre-conditions must be locally true
on the {\em lookup} of the set.
%

In~\cite{shapiro11comprehensive} different CRDT sets are described.
\begin{description}
\item[G-Set] In a Grow Only Set (G-Set), elements can only be added and not
removed. 
\item[2P-Set]
In a Two Phases Set (2P-Set), an element may be added and removed, but
never added again thereafter.
\item[LWW-Set]
In a Last Writer Wins Set (LWW-Set), each element is associated to a
timestamp and a visibility flag.  When two  concurrent operations
 occur, an operation with a higher timestamp is executed.
\begin{figure}[H] 
\centering
\includegraphics[width=8cm]{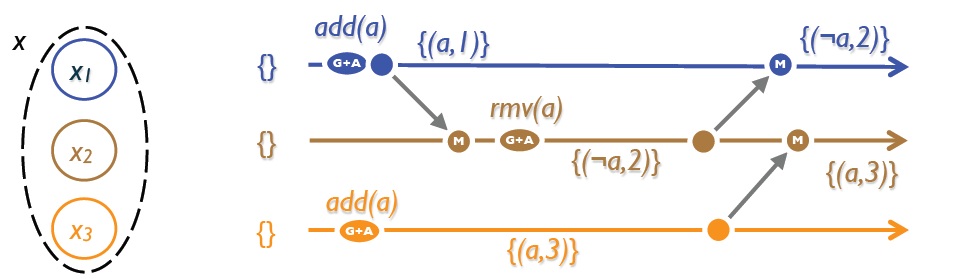} 
\caption{Last Writer Wins Set : LWW-Set~\cite{shapiro11comprehensive}}
\label{fig:LWW} 
\end{figure} 
\item[C-Set]
In a Counter Set (C-Set), each element is associated to a counter. 
 When user add element a counter is incremented, and when user remove an 
 element is decremented. A local $add$ can
occurs only if counter$\leq 0$ and sets the counter to $1$. 
 A local $remove$ can occurs only if counter$> 0$ and sets the counter
to $0$.
\begin{figure}[H] 
\centering
\includegraphics[width=8cm]{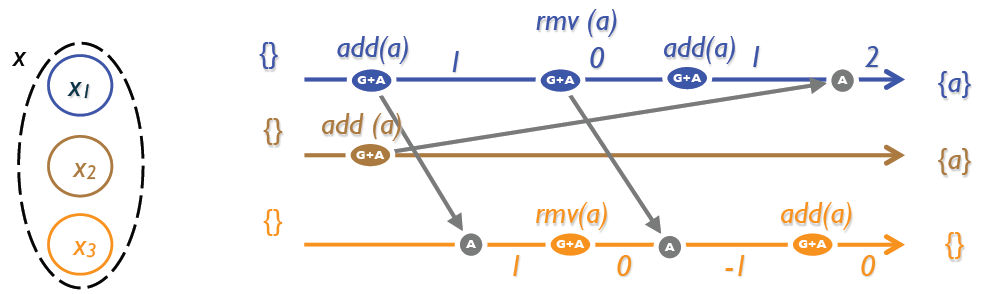} 
\caption{Counter Set : C-Set~\cite{shapiro11comprehensive}}
\label{fig:PN} 
\end{figure} 
\item[OR-Set]
In a Observed Remove Set (OR-Set) each element is represented by
 a unique tag on the set. A local $add$ creates a tag for the element and a local
$remove$ removes all the tag of the element. 
\begin{figure}[H] 
\centering
\includegraphics[width=8cm]{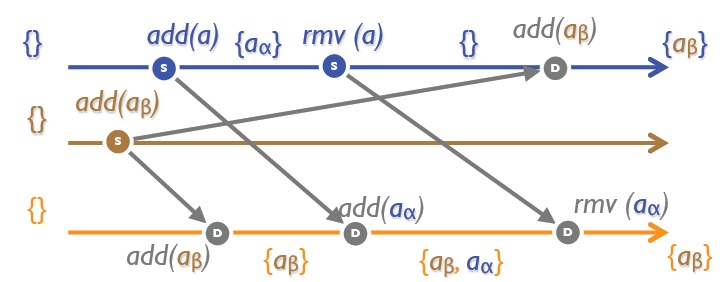} 
\caption{Observed-Remove Set : OR-Set~\cite{shapiro11comprehensive}}
\label{fig:OR} 
\end{figure} 
\end{description}

\section{Layer structure}
\label{sec:layer}

To be able to control and manage conflicts simply, the structure of
the system is managed by layers.  Conflict resolving is invisible to
the user application.  A layer is represented by a component with
the following interfaces:
\begin{description}
 \item[lookup]  the method allows to see the data state; this method represents what users observe.
\item[update]  the method allows to perform modifications on the data. 
\item[replication] the lower layer (and only it) performs
  communication between replica.
\end{description}

Only the lower layer ensure replication and eventual consistency. The
other layers computes a view from the lookup result of their above
layer. Each layer is responsible for a particular constraint :
\begin{description} 
\item[replication] This first layer ensure communication between
  replica and eventual consistency. It ensure the constraint that a
  unique element identified by a path and a type is associated to a
  unique content.  It encapsulates a set CRDT such as described
  previously, Section~\ref{sec:set} and thus, resolves the conflict
  $add(a)||remove(a)$. The encapsulated managed set contains elements,
  i.e. couples $(path, type)$. Beside this set, the replication layer
  maps each file to its content, a CRDT, and resolves the
  $remove(a)||update(b)$ conflict.  The lookup method of the layer
  returns a map $(path, type) \rightarrow content$. For directory, the
  content is empty, the children of a directory is determined using
  paths of other elements. However, a set of path is not a file system
  data structure since other constraints must be ensured.
\item[hierarchical] The second layer is in charge to produce a connected
  tree view from the set of elements provided by the replication
  layer. To produce this tree view, the lookup method of this layer
  has to resolve the conflict $add(a)||remove(b)$ which creates orphan
  nodes. When the update is invoked, it transforms an element in the
  view into a path for updating the set. To obtain this path it must
  take into account how conflicts were resolved by the lookup method.
  To resolve the conflict several type of policies can be defined (see
  Section \ref{sec:file_crdt}). In a tree view returned by this layer,
  a directory may contains several element with the same name (but
  different types and/or different original path).
\item[naming] The third layer ensure uniqueness of element names
  in directories. The lookup method of this layer return the file
  system data structure. In Section \ref{sec:file_crdt}, we present
  two mechanisms to obtain unique names, either by avoiding conflict,
  either by returning a view where original names are changed in case
  of conflict.
\end{description}

\begin{figure}[H] 
\centering
\includegraphics[width=5cm]{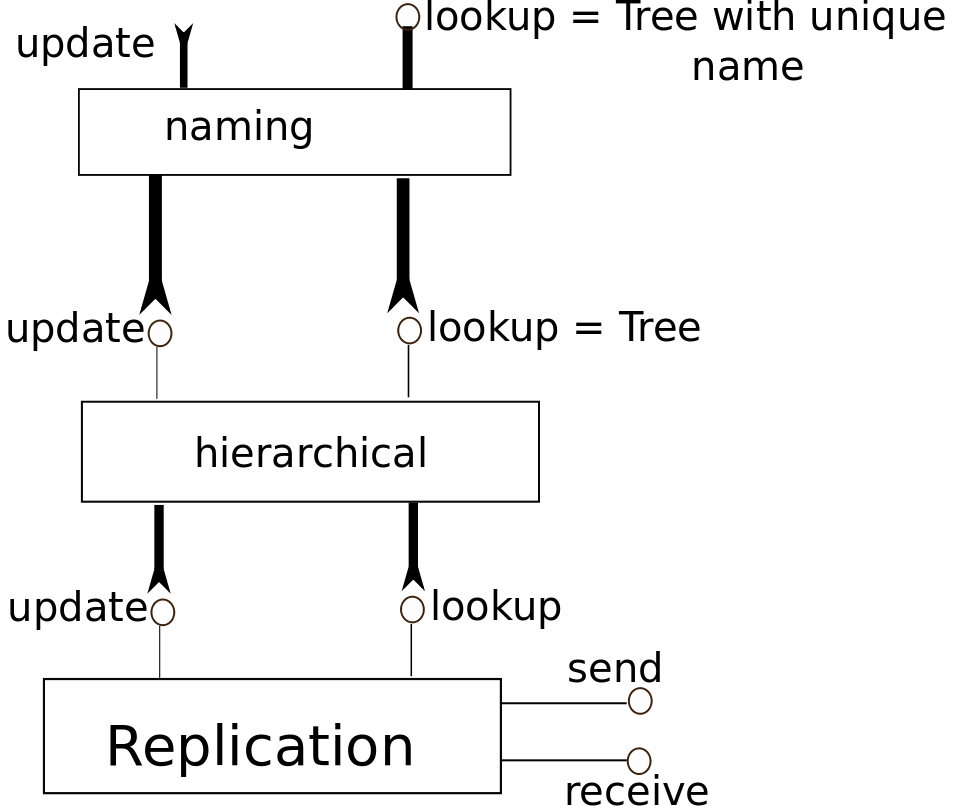}
\caption{Layer structure}
\label{fig:layer_struct} 
\end{figure}

The advantage of this layered management is twofold. First, eventual
consistency is ensured by well-known existing CRDTs. Since the other
layers lookup methods only compute a view, without affecting the inner
replicated state, SEC is ensured. Secondly, such a layered management
allows to combine different solutions for conflict management in order
to obtain a replicated file system. Since each conflict resolution has
its own behavior, and its own computational cost, we give to the
distributed application developer the entire control on the replicated
data structure.


%

\subsection{Replication Layer}

As described above, the replication layer ensure strong eventual
consistency. The update interface of the layer accept three operation
$add(a)$, $remove(a)$ and $update(a, u)$ with $a$ a couple $(path,
type)$ and $u$ and update compatible with the file type of $a$. The
lookup interface return a map $(path, type) \rightarrow content$ with
empty content for directories.

\begin{figure}[H] 
\centering
\includegraphics[width=8cm]{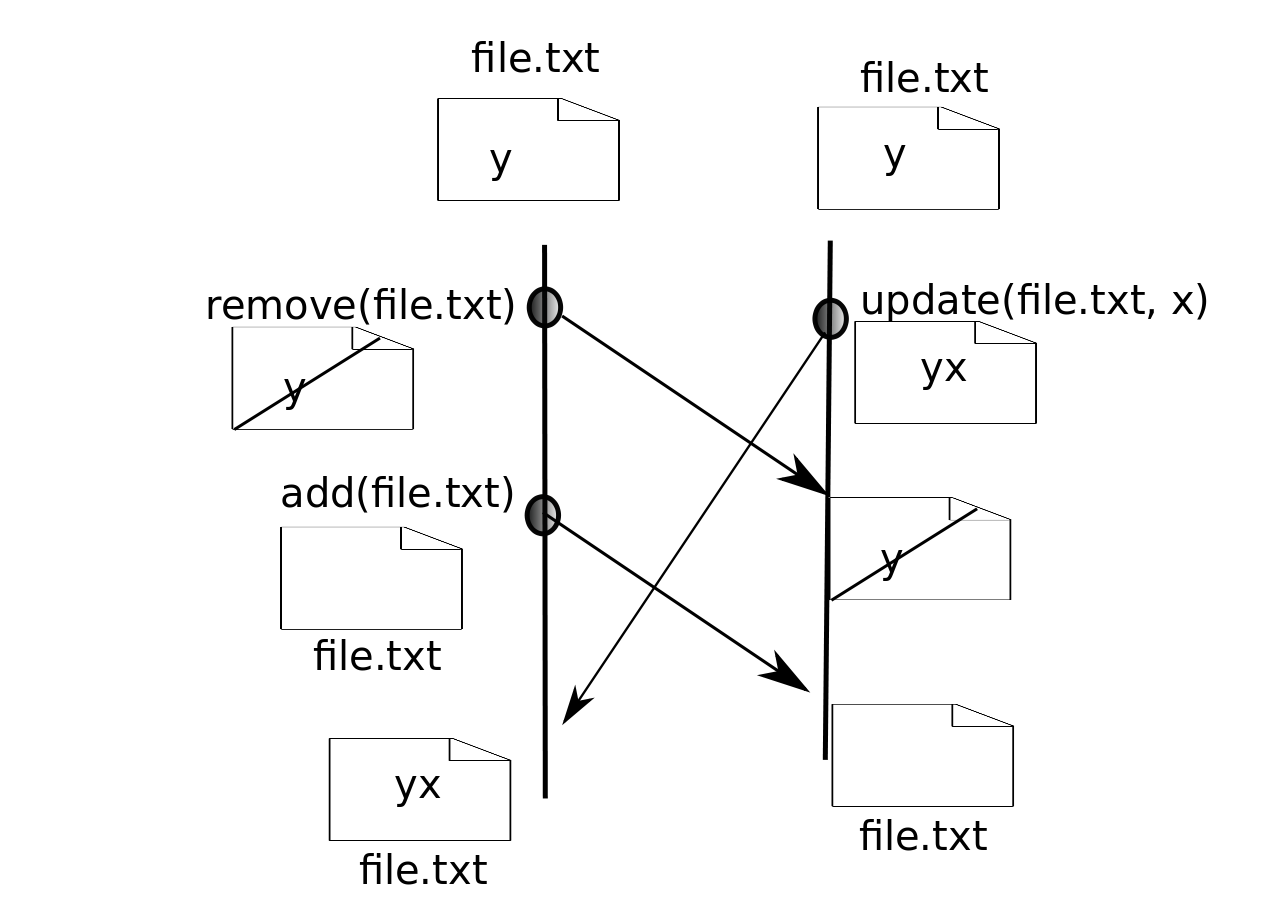}
\caption{Layer structure}
\label{fig:updateremoveadd} 
\end{figure}

The layer encapsulates a set CRDT that contains couples $(path, type)$
to manage the $add(a)||remove(a)$ conflict. Beside this set, the
replication layer maps each element to its content, a CRDT. The layer
keeps content of deleted files. If the content is not kept, the data
would diverge when the element is re-added, See
Figure~\ref{fig:updateremoveadd}. Since the couple $(path, type)$ is
invariant during time, every update is applied on the content of the
element, and eventual consistency of the file content is ensured. This
strategy also resolves the $remove(a)||update(b)$ conflict since both
the file is removed and the content is updated.

However, such tombstone contents should be garbaged somehow. Also, to
ensure the local $add(a)$ post-condition -- the content is empty --,
the local ``add file'' update must creates a couple of operations:
$add(a)$ that makes the file visible and $update(a, u)$ such that
$isEmpty(content(a, S)\circ u)$, i.e. an operation that clears the
file content.

\subsection{Hierarchical Layer}
\label{sec:file_crdt}

This layer is in charge to produce a tree from the set of paths
obtained with the replication layer.  It is in charge to manage the
$add(a)||remove(b)$ conflict where $b$ is an ancestor of $a$.  To
manage this conflicts we propose two kind of solution. The first kind
ignores directory and consider only files, ans thus avoid such a
conflict. The second kind resolve the conflict by treating with
different possible policies the orphan elements that result from such
a conflict.

In both kind of solutions the lookup interface of the layer returns a
tree which labels are tuple $(name, type, path, content)$: the name of
the element with its type and its original path (the path appearing in
the set). The update interface allows to apply the following
operations: $add(p, n, t)$, $remove(p, t)$ and $update(p, t, p', u)$
with $p$ a path in the lookup tree, $n$ a name, $t$ a type, $p'$ the
original path and $u$ a content update.

\subsubsection{Consider only leaf}
\label{sec:tree-leaf}

The first kind of hierarchical layer, consider only the leaf of the
file system. The type ``directory'' no longer appears in the inner
replication layer. To avoid the conflict $add(a)||remove(b)$  we change 
 a pre and post condition described previously (Section \ref{sec:file_sys}).  

 \begin{itemize}
 \item $pre (add(p, n, t), S) \equiv exists(p, S)$ and
   $type(content(p, S))\neq directory$ and $\neg exists(p.n, S)$
 \item $post (add(p, n, t), S) \equiv  exists(p.n, S)$ and
  $type(content(p.n, S))=t$ and $isEmpty(content(p.n, S))$
 \item $pre(remove(p, t), S) \equiv exists(p, S)$.
 \item $post(remove(p, t), S) \equiv \neg exists(p, S)$.
 \item $pre (update(p, t, p', u), S) \equiv exists(p, S)$ and \\ $u$ is
applicable
   on type $content(p, S)$
 \item $post (update(p, t, p', u), S) \equiv exists(p, S)$ and \\ $content(p,
S)' =  content(p, S) \circ u$
 \end{itemize}

This solution has an impact on the inner layer. Indeed, the inner layer
``replication`` contains a set of couple $(p, t)$ with p is a path directed to
files and t is a type such that $type(content(p, S))\neq directory$. Also, the
lookup method of the layer returns a map $(path, type) \rightarrow content
\neq \varnothing$.
 \begin{figure}[H] 
 \centering
 \includegraphics[width=8cm]{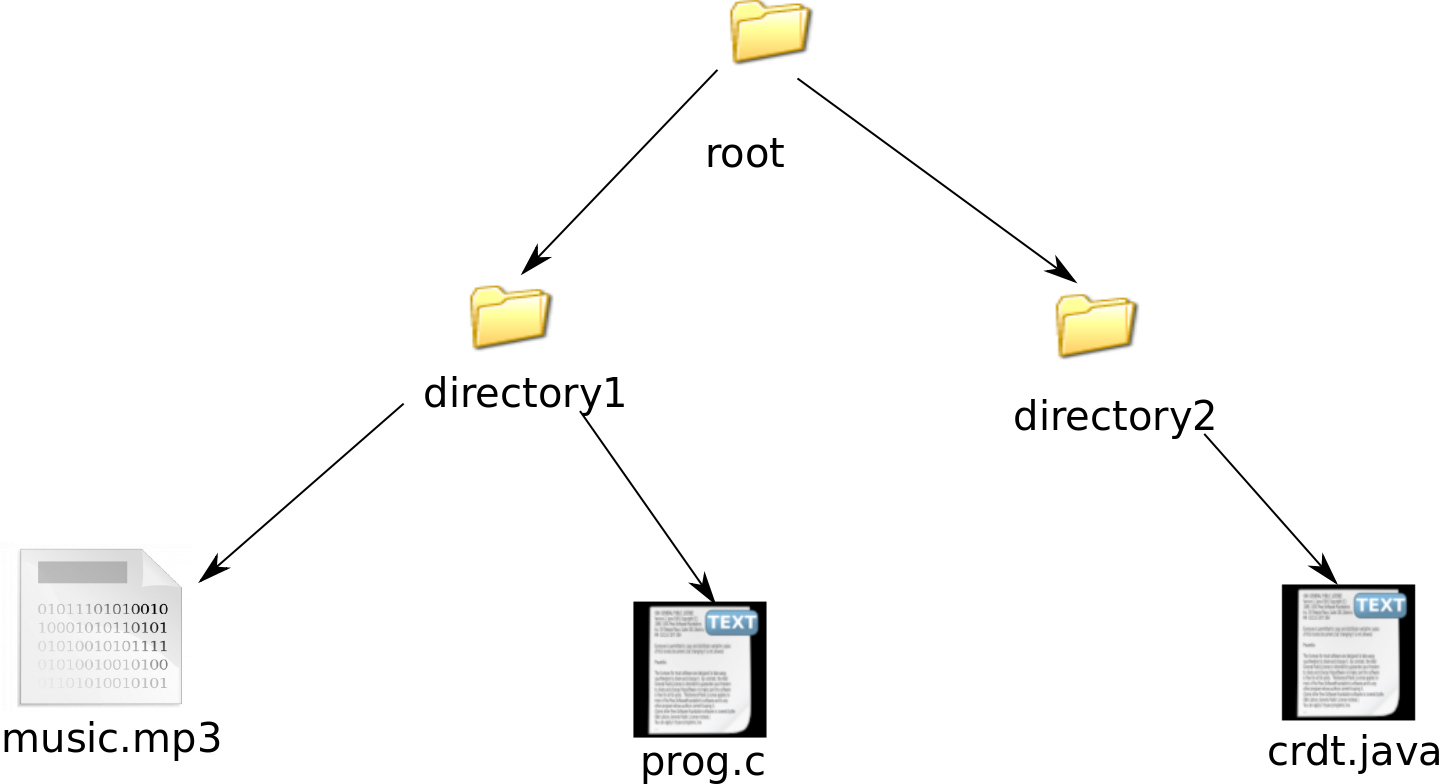} 
 \caption{file system with binary file, text files and directories}
 \label{fig:file_path} 
 \end{figure} 

Exemple :  In the file system respresented in figure \ref{fig:file_path} the
replication layer contains: \\
 $\{(root/directory1/music.mp3, type(music.mp3) = binary ), \\
 ( root/directory1/prog.c, type(prog.c) = text ), \\
 ( root/directory2/crdt.java, type(crdt.java) = text )\}$ \\

 The ``hierarchical`` layer can computes a result tree by using two
methods. First, is an incremental method. In this method, a layer stores the
state of the data type that will be returned to the upper layer and it modifies
this data type each time its inner layer state is modified. The case
of non-incremental method, the lookup recomputes the tree each time its inner
layer state is modified. In both methods, the tree view returned by this layer
is not ready to represente a right file system since a directory may contains
several element with the same name.

The update interface transforms the elements in the view into a path to
invoke an update of the inner layer. 

GIT\cite{git} is based on tree where each file is defined by unique path started
by root. The directories are created on the fly and they represent just a
logical representation to the users. Unlike in our method proposed, in git we
can observe a small divergence. Indeed, if user located in
replica 1 creates an empty directory and commit, a file system as git does not
take into account this empty directory, then, when user 2 makes an update, the
two replica does not observe the same content.

\subsubsection{Treat  orphan nodes}



Another kind of solution to manage the $add(a)||remove(b)$ conflict is
to treat the orphan elements produced by the conflict. An orphan
element is a path in the replication set which its father (its longest
strict prefix) is not in the set. To treat orphans, several policies are
described in \cite{martin11abstract}.
 
The lookup interface obtained by these policies are the following.
\begin{description}
\item[skip] This behaviour does not return orphan element;
  it gives priority to $remove$.  

\begin{figure}[H] 
\centering
\includegraphics[width=14cm]{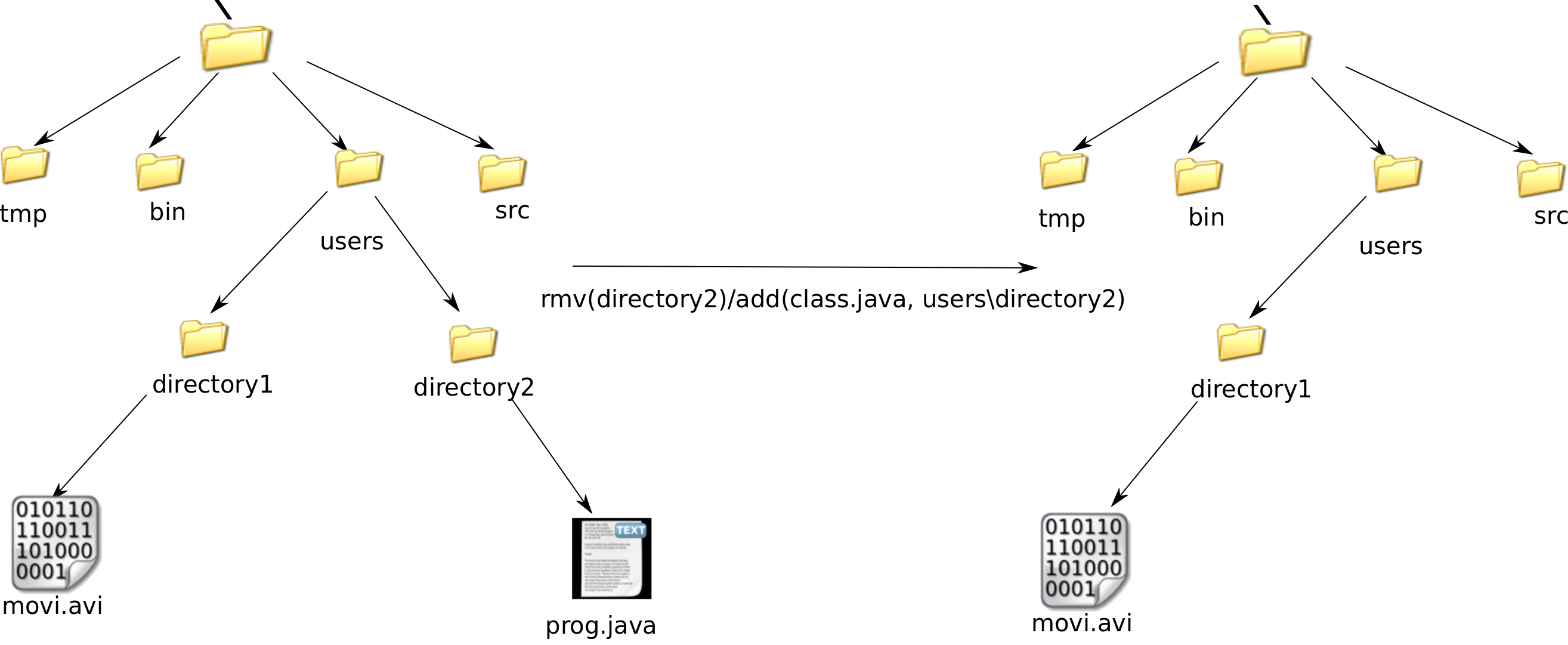} 
\caption{$skip$ policy.}
\label{fig:skip} 
\end{figure} 

\item[reappear] This behaviour returns an orphan element at its
  original path; it give priority to $add$. All required ancestors are
  recreated in the view. However, when recreated directories are empty
  they are removed. This solution has a behavior similar than
  ``\textit{Consider only leaf}'' solution, except than it allows the
  tree to contains empty directories.

\begin{figure}[H] 
\centering
\includegraphics[width=14cm]{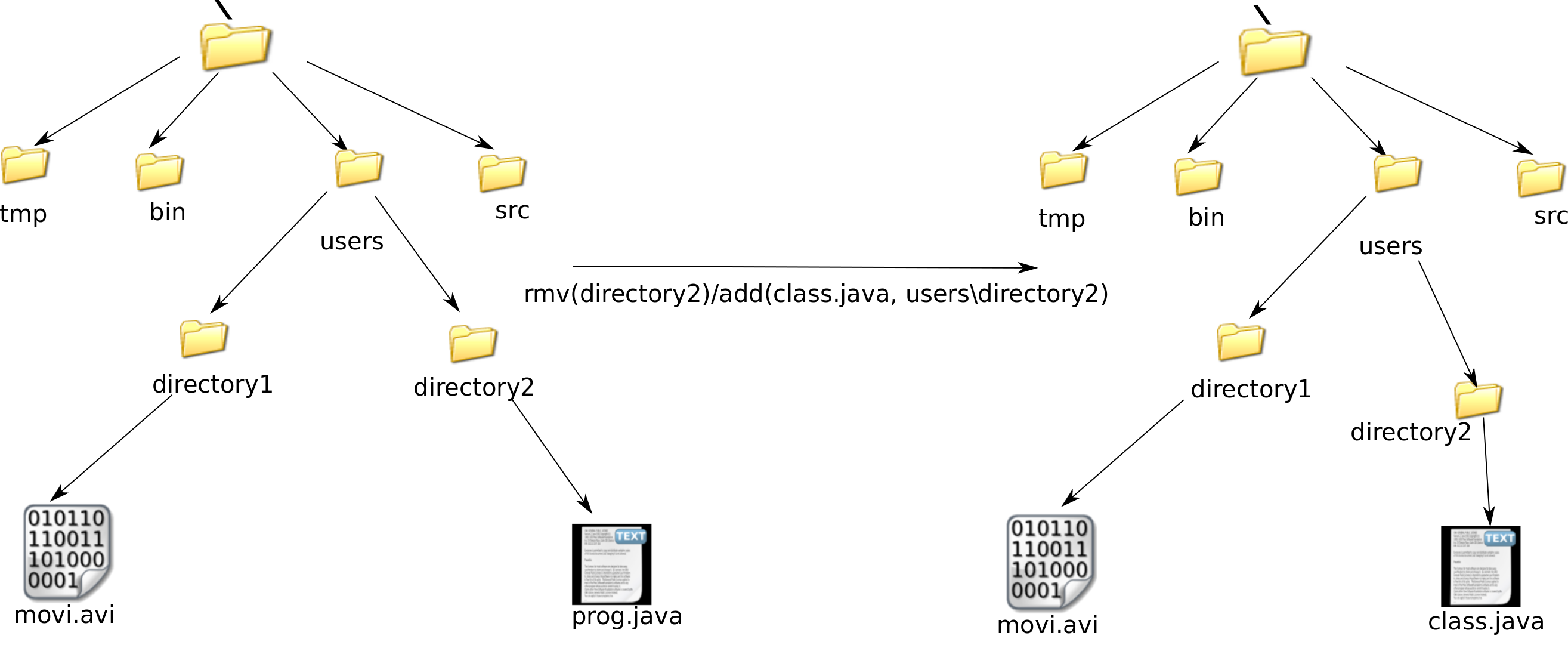} 
\caption{$reappear$ policy.}
\label{fig:reappear} 
\end{figure} 

\item[root] This behaviour places orphan elements under the root.
  This behavior can also be used to place the orphan elements under
  some special ``lost-and-found'' directory as in Ficus replicated
  file system (see Section~\ref{sec:ficus}).

\begin{figure}[H] 
\centering
\includegraphics[width=14cm]{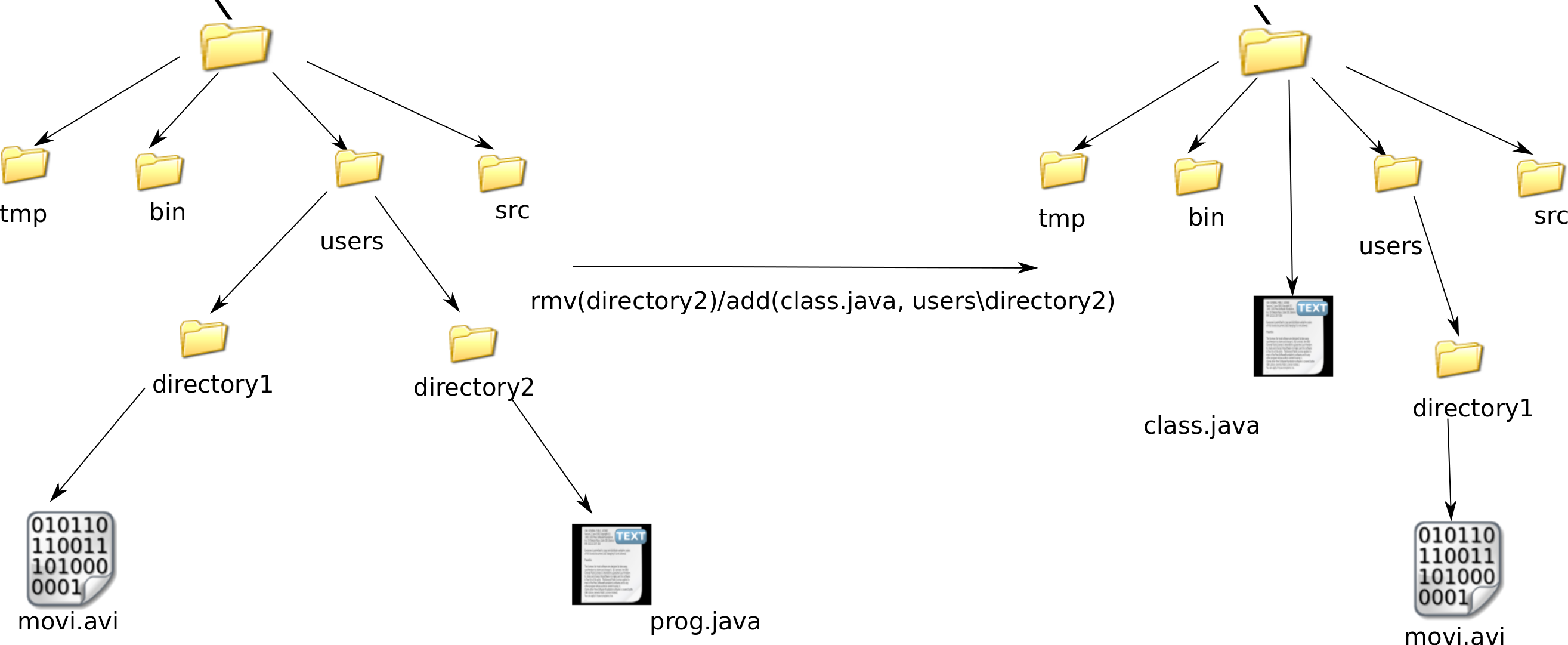} 
\caption{$root$ policy.}
\label{fig:root} 
\end{figure}

\item[compact] This behaviour places an orphan element under the
  longest connected prefix path. In figure~\ref{fig:compact} when the
  file system receives a remove $directory2$ and, concurrently, the
  addition of a file $prog.class$ under $directory2$, this file is
  placed under the father of the deleted directory, i.e. $user$.
 
\begin{figure}[H] 
\centering
\includegraphics[width=14cm]{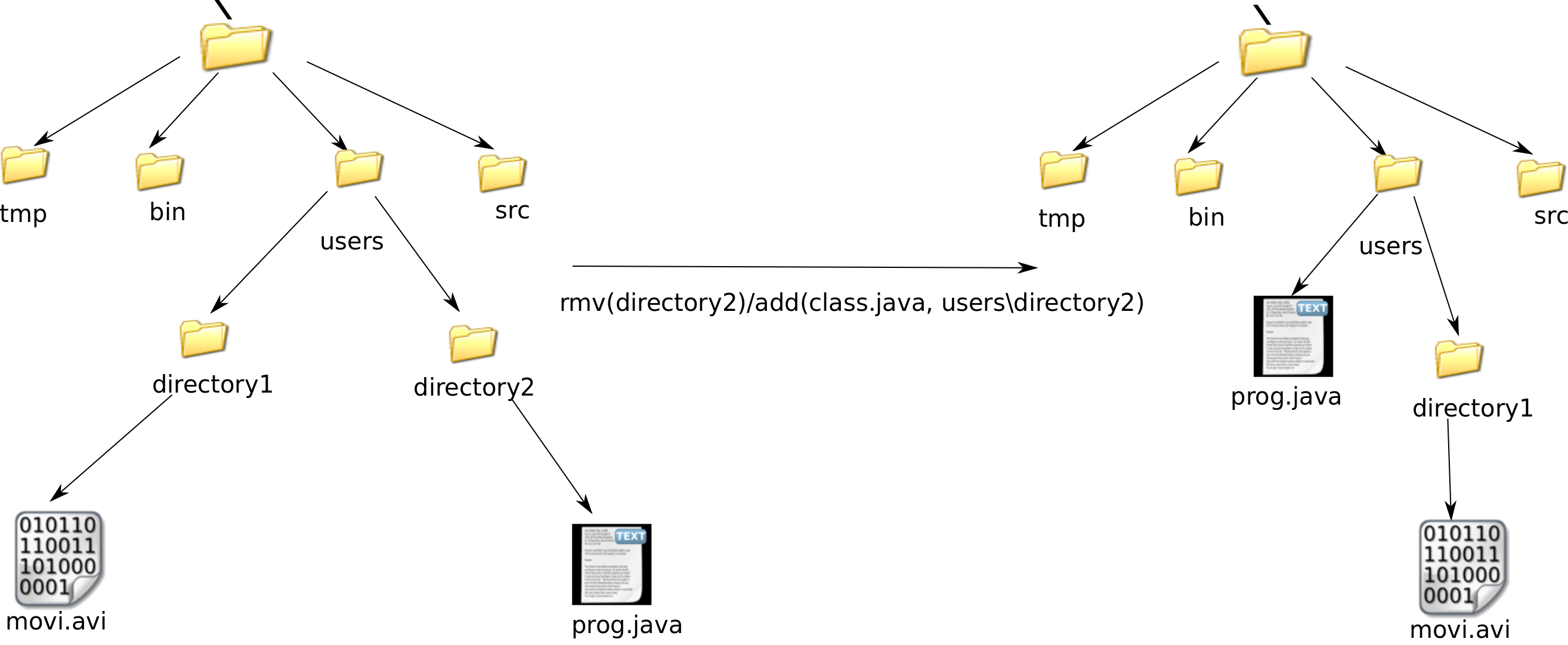} 
\caption{$compact$ policy.}
\label{fig:compact} 
\end{figure}
\end{description}

Each of these policies has a non-incremental version, where the view is
entirely re-computed form the set and an incremental version, where
the view is only updated when a change is made on the inner set. The
update interface adapt path in the tree into path for the set. Due to
choices made by the policies, these paths can be different. For
instance, in Figure~\ref{fig:compact}, $remove(/user/prog.class,
binary)$ will be adapted into $remove(/user/directory2/pog.class,
binary)$. For details, see \cite{martin11abstract}.

Until now, this mechanism returns a hierarchical structure, but it
does not represent yet the file system.  If two replicas add the same
name under the same directory, this structure may confuse. To avoid
this problem and construct a valid file system, we add a layer called
\textit{resolve name} that treats this type of conflict
$add(a)||add(a)$. In what follows we present this layer and how it
treats the conflicts. This layer also treat the case where several
elements with same name and type (but different original paths) where
placed in a directory by the ``root'' or ``compact'' policy.

\subsection{Naming layer}

The naming layer ensures that each directory contains only one
instance one file with a given name. We consider that two files
with the same type created twice concurrently at the same place is
only one file and the content is merged. For elements with different
types or origin, we propose two kind of solutions.

The first method avoid the conflict by altering preconditions of
operation, i.e. it enforces some properties on elements names. The
second method renames on-the-fly conflicting files, so it let
users to resolve the conflict.

\paragraph{First method : }
We associate to each file type a specially algorithm. When
a conflict occur, a file system merge two files since they implements
the same algorithm.

 \begin{itemize}
 \item $pre (add(p, n, t), S) \equiv exists(p, S)$ and
   $\neg exists(p.n, S)$
 \item $pre(remove(p, t), S) \equiv exists(p, S)$.
 \item $pre (update(p, t, p', u), S) \equiv exists(p, S)$
 \end{itemize}

In addition, the directories must not have an extension and text type are not
permitted as an extension for binary files.

 \begin{figure}[H] 
 \centering
 \includegraphics[width=4cm]{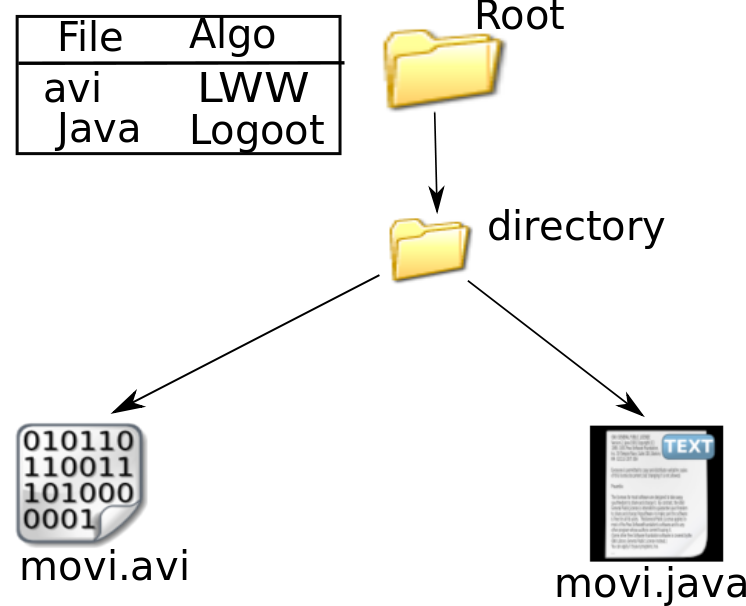} 
 \caption{Different algorithm to each extension}
 \label{fig:StandingAlgorithm} 
 \end{figure}

 In figure\ref{fig:StandingAlgorithm} a tree observed by
user is : $root/directory/movie.java$ and $root/directory/movie.avi$. When
user  makes modification in the file $movie.java$, an algorithm used is
automatically $Logoot$. This method is not permitted for root and compact
policies. Indeed, when a root or compact policies are applied, a file may
located with another under same directory that was not in the same directory
before. In this case, a merge is not permitted since we cannot merge two file
with different origins.

 \begin{figure}[H] 
 \centering
 \includegraphics[width=8cm]{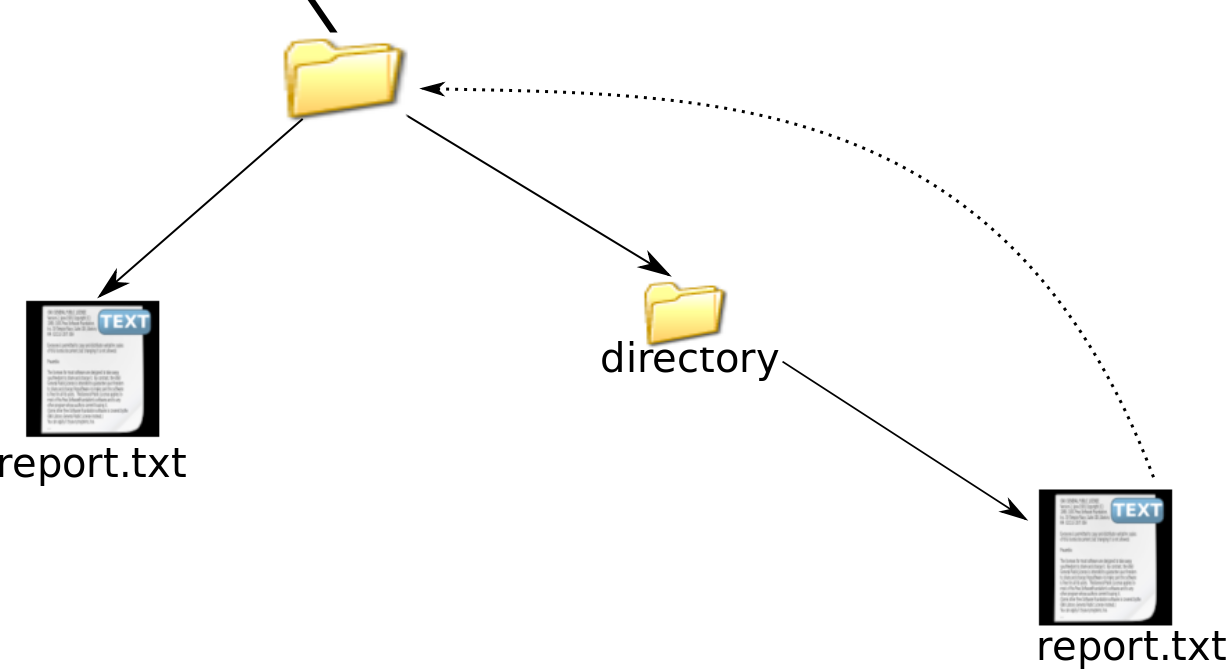} 
 \caption{Two fies conflict after root/compact}
 \label{fig:namingprob1} 
 \end{figure}

\paragraph{Second method : }
 This solution is applied only when conflict occur. To distinguish between
files, we add at the last of file name the name of the algorithm used or the
origin path as an extensions. Finally we propose
to users to choose one of the two files or merge them. In both case we keep
only one file and we remove the extension added from the file name. So, users
can observe a stange behavior of files since it changes name when conflict
disappear (small glish).

In both methods, a lookup interface returns a tree to user application without
conflicts and with unique name. This tree is computed with incremental or 
non-incremental versionss. In case of incremental version, the layer
keep a state of the data structure and the conflict is detected directly when a
method update is invoked. While, in case of non-incremental version,
the layer recomputes all tree each time the users make modifications. The
update interface adapt operations in the tree to detecte and resolve conflicts
names.

\section{Conclusion}
\label{sec:conclusion}

In this report, we have proposed a solution to represent
optimistically replicated file systems. Our solution ensure strong
eventual consistency. We use a CRDT tree to bypass the different
conflicts using a layer structure.  Using a layered approach, each
conflict is managed separately. Thus, we give the choice to the
developer to choose a specific policy to resolve a specific conflict
automatically.  Nonetheless, the final solution concerning unique
names have some drawbacks, first, it changes a name of files in case
of conflict which is not desirable to users, and second the conflict
in some cases is resolved manually by users. However, this method
gives more alternatives to developers compared to other methods.
   
Finally, our approach produce a best effort merge that may satisfy the
application developer but not all the final user of the
application. So such solution have to be coupled to an awareness
mechanism~\cite{dourish92awarness} that allows the user to be
conscious of the choice made automatically by the system and to
produce updates that correspond to another choice.

\bibliographystyle{abbrv}
\bibliography{theBib}

\end{document}